\documentstyle[11pt,newpasp,twoside]{article}
\def\edcomment#1{\iffalse\marginpar{\raggedright\sl#1\/}\else\relax\fi}
\reversemarginpar

\begin{document}

\title{The vertical structure and kinematics of HI in spiral galaxies}
\author{Renzo Sancisi}
\affil{Astronomical Observatory, Bologna, It \& Kapteyn Inst., Groningen, NL}
\author{Filippo Fraternali}
\affil{I.R.A.(CNR) \& Astronomy Department, Bologna, It}
\author{Tom Oosterloo}
\affil{N.F.R.A., Dwingeloo, NL}
\author{Gustaaf van Moorsel}
\affil{N.R.A.O., Socorro, NM, USA}

\begin{abstract}
{We discuss the distribution and kinematics of neutral hydrogen in the
halos of spiral galaxies.
We focus in particular on new results obtained for the nearby Sc galaxy 
NGC~2403  which have revealed the presence of extended HI emission 
(here referred to as the `beard')
at anomalous velocities with respect to the `cold' HI disk.
Modeling shows that this component has a mass of about 
1/10 of the total HI mass and is probably located in the halo region. 
Its kinematics differs from that of the thin HI disk: it rotates more slowly 
and shows radial inflow. The origin of this anomalous gas component is 
unknown. It could be the result of a galactic 
fountain or of accretion of extragalactic `primordial' gas.}
\end{abstract}

\section{Introduction}
The study of the vertical structure and kinematics of the HI disks of 
spiral galaxies has in recent years been pursued in two main directions.
One has been the investigation of the large-scale dynamics, with special 
attention to the phenomena of warping and flaring in the outer parts 
of the HI disks. Attempts have been made (e.g. Van der Kruit 1981, 
Olling 1995) to determine the mass of the stellar disks and to 
investigate the properties of dark matter halos.
The other has been the search, in the inner parts of the disks, 
for evidence of vertical extensions and motions of the HI gas, possibly 
related to star formation activity, the expansion of superbubbles and 
the circulation of gas between disk and halo. 
The results are helpful for understanding the 
observations of the Milky Way and indeed, as we will see 
below, some of the HI features discussed here do have 
properties similar to those of the galactic High Velocity Clouds.

In addition to these `internal' phenomena, there are 
other events of external origin. These are the gravitational interactions 
with other systems and the accretion of gas, which may affect the density 
distribution and kinematics of the gas in the halo region and may also 
play an important role in the formation and evolution of 
disks (see for a recent review Sancisi 1999).

\section{Edge-on and face-on galaxies}
Traditionally, the study of the disk-halo connection and gas 
circulation has been carried out on edge-on and face-on systems, 
the former to obtain the projected distribution of the HI density, 
and the latter to observe possible vertical motions.
Probably the best studied edge-on galaxy is NGC~891.
The HI maps obtained 
by Rupen (1991) and by Swaters, Sancisi, \& van der Hulst
(1997) show the presence of emission extending up to about 5 kpc
on both sides of the plane of the galaxy.
This emission had been interpreted in the early study of Sancisi \& Allen 
(1979) as a huge outer flare of the gas layer seen in projection. Later on 
Becquaert \& Combes (1997) proposed a line-of-sight warp.
More recently, Swaters {\it et al.} (1997) have carried out a detailed 3D 
analysis of the data investigating the possible projection effects.
They have concluded that the extended emission comes from gas 
in the inner halo region of NGC~891 (and not in the outer flare) and have 
proposed that this gas has a rotation velocity about 25 to 100 km~s$^{-1}$ 
lower than the gas in the plane. They have tentatively attributed this velocity 
difference to a gradient in the gravitational potential and have pointed 
out that this may serve to discriminate between disk and spheroidal mass 
distributions.  

The HI in the halo of NGC~891 has a mass of about 6$\times$ 10$^8$ 
M$_{\odot}$, corresponding to 15\% of the total HI mass and to 
$\sim$0.3\% of the total dynamical mass.
The origin of this halo gas is probably related to star formation 
activity in the disk of NGC~891 and to disk-halo circulation.
This is supported by the presence of a thick disk component which is 
observed in the radio continuum (Dahlem, Dettmar, \& Hummel 1994) and 
in H$\alpha$ (Rand, Kulkarni, \& Hester 1990), and is also indicated by 
the extended dust structures visible in the high-resolution optical 
WIYN images (Howk \& Savage 1997).
The lack of information on the vertical motions is obviously the main 
limitation in the study of edge-on galaxies like NGC~891.

The motion of the gas in the vertical direction 
has been studied in several nearly face-on spiral 
galaxies. Emission with non-zero vertical velocity has been found, 
often associated with large `holes' in the cold gas distribution.
Detailed studies have been done, for example, in M33 (Deul \& den 
Hartog 1990), M101 (Kampuis 1993) and some dwarf galaxies like HoII 
(Puche {\it et al.} 1992).
The general interpretation of these features is that they are the 
results of the formation and expansion of supershells around stellar  
associations due to stellar winds from massive O and B stars and/or 
supernova explosions.
A link between the vertical motions of HI and star formation activity 
is also supported by the high values of the velocity dispersion of HI 
found in the central, bright optical regions of these galaxies 
(Kamphuis 1993).
Such HI observations of face-on galaxies do not provide, however, any 
direct information on the z-distribution of the gas.

\begin{figure}[ht]
\plotfiddle{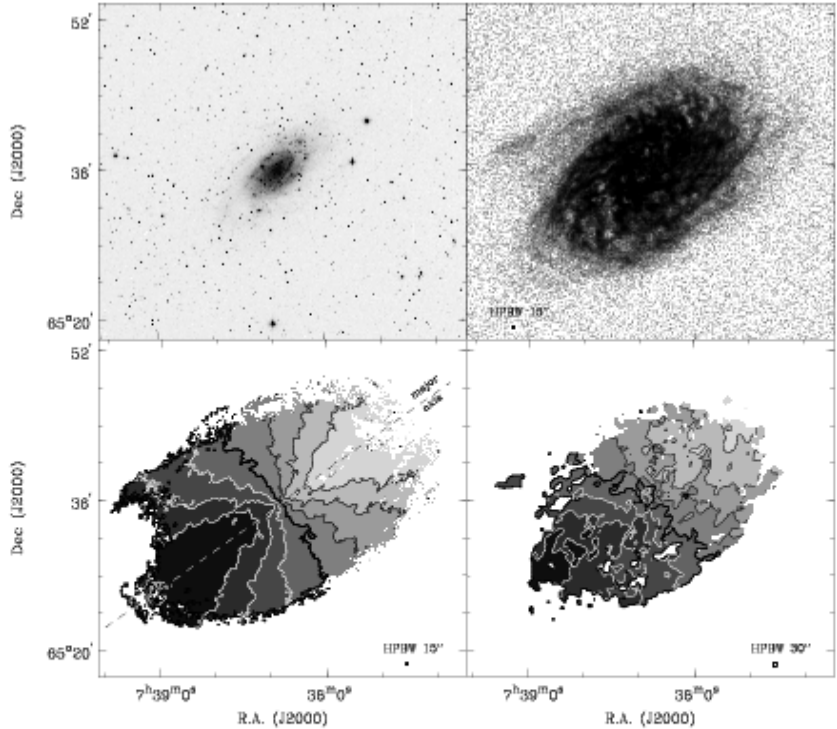}{9.9cm}{0}{90}{90}{-220}{-90}
\caption{\small NGC~2403. Upper panels: optical image and HI density map 
(VLA). 
Bottom panels: velocity field of all HI gas (left) and
of the ``anomalous'' gas only (right). All plots are on the same scale, 
1 arcmin $\simeq$ 1 kpc.}
\label{fig1}
\end{figure}

\section{Galaxies with `beards'}
The limitations and difficulties in studies of the vertical
structure of HI that are encountered with galaxies viewed
either edge-on or face-on can be partly overcome by taking objects of
intermediate inclinations.
The nearby spiral galaxy NGC~2403 (see Fig.~1) offers a good illustration 
and is an excellent candidate for such a study because of its ideal 
inclination of $\sim$60$^{\circ}$, its extended HI layer (about
twice the optical), its regular kinematics and symmetric, flat rotation
curve.

The position-velocity diagram along the major axis of NGC~2403 shows 
systematic asymmetries in the form of wings on the side of the
lower velocities, towards the systemic velocity.
This was already visible in the early WSRT data (Wevers, Van der Kruit, \& 
Allen 1986, p.~551, Fig.~6a, Begeman 1987, p.~46, Fig.~3) but it is 
more clearly shown in our new VLA data (see Fig.~2, right panels). 
Such a pattern (we refer to it as the `beard') is similar to that found 
in edge-on galaxies and also in objects observed at relatively low 
angular resolution. Both VLA and WSRT observations of NGC~2403 have, 
however, a sufficiently high angular resolution and the galaxy is not 
too highly inclined. It is, therefore, quite surprising to see such 
systematic asymmetries in the position-velocity maps.

What can be the explanation for the beard? Clearly, it cannot be simply 
the result of a high velocity dispersion or of gas moving away from 
or toward the disk in the perpendicular direction, because that would produce 
extensions symmetric with respect to the rotation velocity.

Recently Schaap, Sancisi, \& Swaters (2000) have analyzed the WSRT
observations obtained by Sicking (1997) and have investigated the 
effect of the layer thickness on the shape
of the HI velocity profiles along the major axis of NGC~2403.
In order to do this they have constructed 3D models of the galaxy, examining 
the possibilities of $i)$ a thick HI disk, as opposed to the usually assumed 
thin layer, and $ii)$ a two-component structure, with a thin layer and a 
thicker but less dense one.  
They have concluded that, in order to reproduce the
observations, the HI disk of the one-component model considered in $i)$
would have to be unrealistically thick, with a FWHM of at least 5 kpc.
The best solution has been obtained by considering a model as $ii)$ with a 
slower rotation velocity for the gas located above the plane of the disk.
They found for this `halo' component a mass of about 15\% of the total HI mass and
a difference in rotation velocity of about 25-50 km~s$^{-1}$.
Clearly, this is a result very similar to that found by
Swaters {\it et al.} (1997) for NGC~891.

\subsection{The VLA Observations}
We present here the results of a new investigation of the HI structure and 
kinematics of NGC~2403 with observations obtained with the VLA in C 
configuration and 48 hours integration.  
Fig.~1 shows the optical picture, the total HI map (top right) and 
the HI velocity field (bottom left).
These data have significantly better sensitivity and higher velocity resolution 
than those used by Schaap {\it et al.} (2000) and have led to a
considerable improvement of the observational picture.  
The existence of the `beard' is confirmed and also new surprising 
features have been discovered.

Fig.~2 shows three representative channel maps and three position-velocity 
diagrams for NGC~2403. 
The position-velocity diagrams are taken
along the major axis at position angle 124$^{\circ}$ (the central panel) and parallel 
to it at $+$~2$'$ and $-$~2$'$ from the centre of the galaxy.
The slice along the major axis shows pronounced tails towards the 
systemic velocity (the `beard') and also unexpected emission in the quadrants
of `counterrotating' or 'forbidden' velocities.
These features are recognizable also in the other slices.
Their spatial location can be seen in the channel maps.
The upper map clearly shows the location of the `forbidden' gas in the inner
($\sim$4 kpc) region of the HI disk, to the North-West of the centre.
The bottom plot shows the faint `forbidden' emission on the South-East side of
the galaxy and the central panel shows a remarkable 8 kpc long HI filament.

Also on the high velocity side of the line profiles there are some
weak extensions of HI emission up to about 25-30 km~s$^{-1}$ with respect to
the rotational velocity which may represent de-projected vertical motions
of about 50-60 km~s$^{-1}$. These are, however, much smaller deviations 
than those of the extended `beard' and 
of the `forbidden' gas which has projected differences from the rotation
velocity of up to 150 km~s$^{-1}$.

\begin{figure}[ht]
\plotfiddle{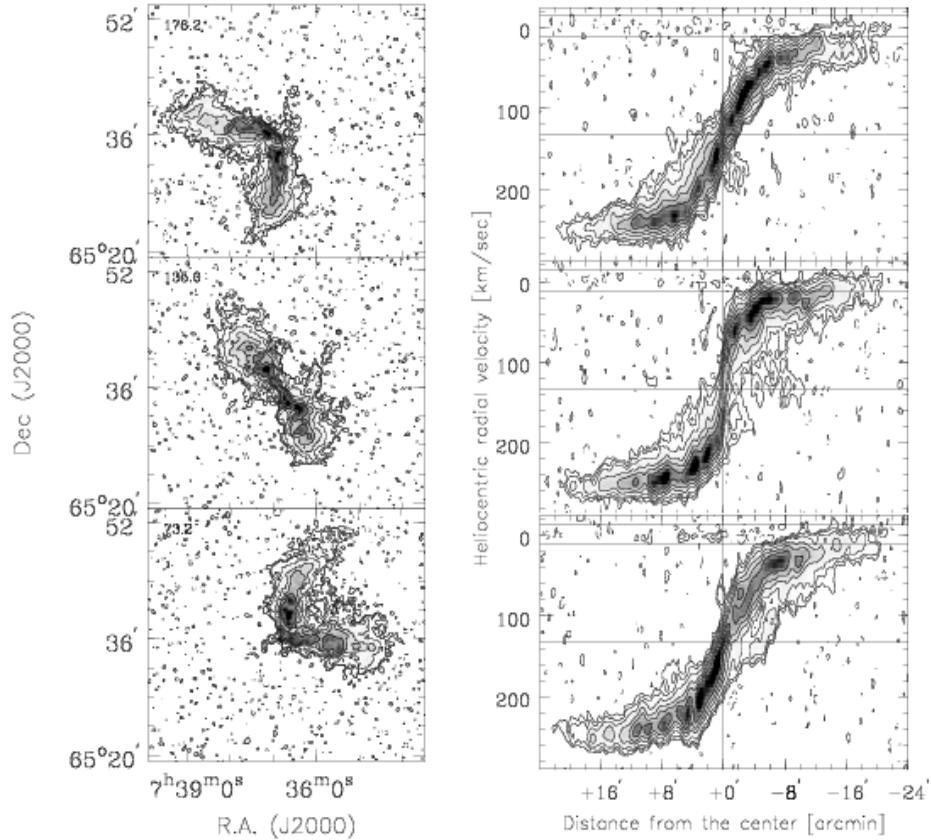}{11.6cm}{0}{80}{80}{-220}{-105}
\caption{\small 21-cm line observations of NGC~2403 with the VLA. 
	Left: three representative channel maps. The centre of NGC~2403 is
	indicated by the cross. 
	Right: position-velocity map of HI along the major axis of 
	NGC~2403 (PA=124$^{\circ}$) and, top and bottom panels, parallel to the
	major axis (centred at the minor axis at 
	positions +2$'$ (N-E) and -2$'$ (S-W) from the centre).
       Contours are: -0.5, 1, 2, 5, 10, 20, 50 mJy/beam; r.m.s. noise
is 0.22 mJy/beam.} 
\label{fig2}
\end{figure}

The overall picture that emerges from the new data is that
of an extended zone of gas which seems to `know' about the general pattern of
disk rotation but deviates from it strongly.
This gas is detected out to about 15 kpc from the centre of the galaxy (the
radius of the total HI disk is more than 20 kpc). The velocity deviations 
are largest in the inner regions. 
We will refer to all this gas detected at velocities significantly different 
from rotation as the `anomalous' gas.
Its total mass is 3.5$\times$10$^8$ M$_{\odot}$, which corresponds to
$\sim$10\%  of the total HI mass of NGC~2403, and to $\sim$0.3\% of its total 
dynamical mass.
Similar values were found for NGC~891 (Swaters {\it et al.} 1997).
The spatial structure of this gas shows a certain clumpyness with filaments, 
up to $\sim$10 kpc in length, and spurs with masses of up to 10$^7$ 
M$_{\odot}$ as seen in the channel maps of Fig.~2.

\subsection{Models}
The first step in the analysis of the HI observations reported above 
has been the construction of 3D models for the gas layer of 
NGC~2403 in order to test simple possibilities such as that of a single 
thick gas layer, of an outer flare and of a line-of-sight warp. 
We have been able to exclude them and to conclude, in agreement with  
Schaap {\it et al.} (2000), that such anomalous emission must come 
from a component with peculiar kinematics.
In order to determine the properties of such a component we
have proceeded to separate the anomalous gas from the cold, thin HI disk.
This has been done by assuming that the line emission from the thin disk 
can be represented by a 
Gaussian profile symmetrical with respect to the rotation velocity. 
We have fitted such profiles at each position and subtracted them 
from the data. 

Fig.~1 (bottom right) shows the resulting velocity field for the 
anomalous gas, which can now be compared with that derived for the total 
HI (bottom left).
The kinematics of the anomalous gas is clearly dominated by differential 
rotation, but the mean rotation velocities are lower than in the regular disk, 
about 50 km~s$^{-1}$ in the inner and 20 km~s$^{-1}$ in the outer parts.  
Moreover, its kinematical minor and major axes appear to be somewhat rotated 
as compared to those of the regular disk and are non-orthogonal.
A likely explanation for this may be radial in-flow of gas towards 
the centre of the galaxy.

\begin{figure}[ht]
\plotfiddle{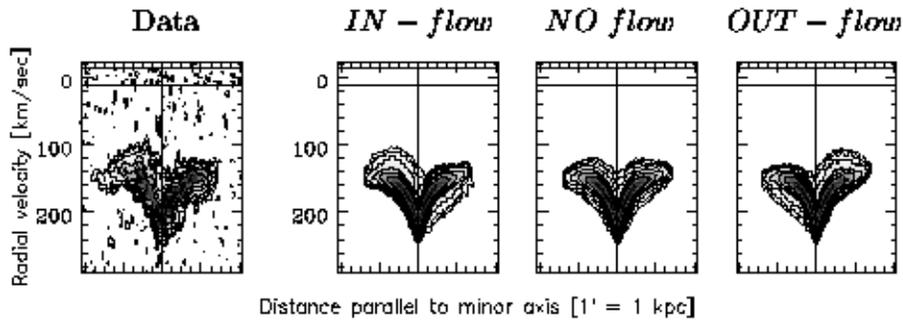}{4cm}{0}{80}{80}{-240}{-265}
\caption{\small Left: position-velocity diagram parallel to the minor
axis 
of NGC~2403 centred on the major axis at 2$'$ S-E from 
the centre of the galaxy. 
Right: three models, characterized by radial in-flow, no flow and
out-flow.}
\label{fig2}
\end{figure}

In order to test this hypothesis, we have constructed 3D models 
of the HI layer of NGC~2403 adopting a two-component structure: $-$a thin cold
disk and a thicker layer rotating more slowly$-$.
Fig.~3 shows a position-velocity diagram parallel to the minor 
axis at 2$'$ (South-East) from the centre of the galaxy, a position 
suitable to illustrate the effects of radial motions.
The diagram shows asymmetries (in the ``V'' shape) especially visible at 
the low density levels which are produced by the anomalous gas component.
For the thicker layer (or halo) component we have adopted a smaller
rotation velocity, a lower density and a higher velocity dispersion as
compared to the thin disk component. 
All these quantities (except the velocity dispersion assumed 20 km~s$^{-1}$)
are derived from the data after the separation between the
two components.
The two models labeled with ``in-flow'' and ``out-flow'' are obtained by 
adding, for the anomalous component, a constant radial motion of $-$20 
km~s$^{-1}$ and $+$20 km~s$^{-1}$ respectively.
The anomalous HI in NGC~2403 clearly indicates a preference for in-flow.

Configurations involving a mis-alignment of
the rotation axes of the halo and of the disk, as proposed in connection with
the origin of warps (Debattista \& Sellwood 1999), have also been considered
here for NGC~2403. A mis-alignment of the axes with no distorsions in the 
halo potential is likely to produce a change of the position
angle of major and minor axis but not the observed non-orthogonality of 
the two. This seems to argue in favour of radial motions.
Another way to explain the non-orthogonality of the kinematical axes would be to
assume a pattern of elliptical orbits as expected for a triaxial halo 
potential. 
Such a possibility is, however, unlikely and not supported by 
the harmonic analysis of the velocity field of NGC~2403 made by Schoenmakers,
Franx, \& de Zeeuw (1997).  

Finally, the decrease in rotation velocity and the high velocity dispersion 
could be the result of 
a large asymmetric drift of an HI halo. This possibility has not been 
investigated yet.

\section{Discussion and Conclusions}
What are the origin and nature of the anomalous gas in 
NGC~2403?
The overall pattern and the large-scale regularity indicated by the 
position-velocity maps (Fig.~2) suggest that this gas (both the `beard' 
and the forbidden emission) forms one coherent structure. 
Its anomalous velocities, despite a big spread of more than 150 km~s$^{-1}$,
are confined inside the range of the rotational velocities of 
the disk. 

The overall pattern of this anomalous gas and its coherent spatial and 
velocity structure seem to point to one phenomenon and one common origin. 
Possibilities are:

1. A galactic fountain (Shapiro \& Field 1976; Bregman 1980) powered by
supernova explosions and stellar winds which blow up the ionized gas from the
disk into the halo. The gas may reach large distances from the plane and then
cool and fall back onto the plane.
This explanation finds some support in the active star formation and the 
presence of bright HII
complexes in the disk of NGC~2403. Also the presence of a large number of
holes in the HI layer (see Fig.~1) and of expanding supershells 
(Mashchenko, Thilker, \& Braun 1999) supports the picture of an effervescent 
disk and of a significant disk-halo-disk circulation.
We are pursuing the study of these phenomena further with deep spectroscopy 
of the H$\alpha$ emission from the bright inner
parts of NGC~2403 and $Chandra$ observations of the bright optical disk.
The main difficulty with a standard fountain interpretation
lies in the presence of the apparently counterrotating, `forbidden' emission
in NGC~2403.
In order to explain that, a new approach and different assumptions for the
fountain dynamics, for instance no conservation of angular momentum, may be
necessary.

2. Infall of {\it primordial} intergalactic gas. The hypothesis of 
cosmological gas infall has been proposed in the past and discussed 
(cf. Oort 1970) in connection with the problem of the origin of the HVCs 
in our galaxy and with the need to re-supply the disks of spiral galaxies with
fresh gas. Such a possibility has to be considered here. In that connection it
would be essential to understand the overall kinematical pattern of the
anomalous HI and to understand its large-scale coherent
structure which follows the rotation of the disk of NGC~2403.  Also, it would
be important to determine its metal abundance and to find out whether it is as
low as found for the intergalactic gas or closer to solar.
 
The results of the present HI observations of NGC~2403 have also interesting
implications regarding the High Velocity Clouds observed in the Milky Way
(Wakker \& van Woerden 1997) which are still a mystery and whose distances,
nature and origin are still a matter of debate.  
Is it possible that the anomalous HI complexes discovered around NGC~2403 
are analogous to the HVCs seen in the Galaxy? 
With what we know of the HVCs it seems fair to suggest
that we probably have detected in NGC~2403 a population of HI clouds of the
same type as those intermediate and high velocity
clouds which are thought to be closely associated with the Galaxy and to be
probably a galactic fountain type of phenomenon.  Clearly, they would be
different from the Magellanic Stream.

How common are the `beards' of galaxies? As far as we know there are some
cases, like M~33 and UGC~7766, which
indicate the presence of similar structures. HI
observations of higher sensitivity 
would tell whether there is any correlation with 
luminosity or mass or high surface brightness or
environment, and which of these may play the crucial role.

\end{document}